# Atomic mechanisms of fast diffusion of large atoms in Germanium

Yuhan Ye[1+], Manting Gui[1+], and Jun-Wei Luo[1,2]*

[1]*College of Materials Science and Opto-Electronic Technology, University of Chinese Academy of Sciences, Beijing, China*

[2]*State Key Laboratory for Superlattices and Microstructures, Institute of Semiconductors, Chinese Academy of Sciences, PO Box 912, Beijing 100083, China*

+These authors contributed equally to this work.
*Email: jwluo@semi.ac.cn

## Abstract

The performance of strained silicon as the channel material for transistors has plateaued. Motivated by increasing charge-carrier mobility within the device channel to improving transistor performance, germanium (Ge) is considered as an attractive option as a silicon replacement because it has the highest p-type mobility of all of the known semiconductor materials and to be compatible with today's conventional CMOS manufacturing process. However, the intrinsically high carrier mobility of germanium becomes significantly degraded due to, unlike silicon, germanium's native oxide is unstable and readily decomposes into several GeyOx suboxides with a high density of dangling bonds at the interface. In addition, these interface trap states will also degrade the off-state leakage current and subthreshold turn-off of a germanium-based device, significantly affecting device scalability. Furthermore, obtaining low-resistance Ohmic contacts to n-type Ge is another key challenge in developing Ge CMOS. To solve these challenges, extensive efforts have been made to attempt the incorporation of new materials, such as Al2O3, SiN3, TiO2, ZnO, Ge3N4, MgO, HfO2, SrTiO3, and Y2O3, into Ge transistors. Control of the diffusion of foreign atoms into Ge is therefore a critical issue in developing Ge transistors regarding foreign impurities may be detrimental for devices. In this work, we study the diffusion properties (or energy potential barriers along the migration pathway) of all common elements in Ge by performing the first-principles calculations with a nudged elastic band method. We find some large atoms, such as Cu, Au, Pd, etc., have a very small diffusion energy barriers. We reveal the underlying mechanism in a combination of local distortion effect and bonding effect that controls the diffusion behaviors of different atoms in germanium. This comprehensive study and relatively in-depth understanding of diffusion in germanium provides us with a practical guide for utilizing germanium more efficiently in semiconductor devices.

# I. INTRODUCTION

Although the very first transistor was not made from silicon but built on germanium (Ge) substrate, silicon and its oxide became the major material in semiconductor device field later [1]. The development of Ge-based devices was hindered by the instability of a stable native germanium oxide which is essential for gate dielectric and by the lack of detailed research about doping properties in Ge [2]. But Ge still has many advantages like higher charge carrier mobility than silicon, narrower band-gap and a generally lower doping activation energy. The compatibility with Si-based devices processing technology makes study on Ge even more significant [1][3]. Particularly, it makes Ge one of the most suitable materials for epitaxial III-V growth substrate due to its very small lattice mismatch with GaAs [1].

Doping is one of the most feasible processing technologies to prepare semiconductors into the desired properties [4]. Based on the various diffusion behaviors of different elements, however, they have different limits and requirements. While bulk doping needs high diffusion rate and small diffusion barriers, slow diffusion rate and relatively large migration energy are more proper for surface modification. Many technologies have been developed utilizing special diffusion behaviors in Ge, like metal induced lateral crystallization (MILC) [4][5][6], in which Ge is grown on palladium (Pd) crystallization-initializing thin film. It is realized based on the ultrafast diffusion of Pd in Ge. Despite the significance of diffusion behaviors in Ge for the industrial modification, it has not been fully understood. Thus, the precise control of dopant diffusion required for Ge-based devices would be greatly aided by an accurate understanding of various impurity diffusion properties in Ge [7].

Typically, there are more than one mechanism controlling the diffusion behaviors in Ge [1][5]. Direct interstitial diffusion is one of the diffusion mechanisms in Ge result in smallest energy barrier. The atom residing in a tetrahedral interstitial site will diffuse to a neighboring tetrahedral site along a straight pathway, passing the hexagonal interstitial site [5][8]. Many metal atoms like Cu, Pd are known to diffuse in this way in Ge. Besides, the vacancy mediated mechanism is pervaded for diffusion in Ge, where the interaction between donor atoms and the vacancy controls the diffusion [9]. The dopant jumps to a neighboring intrinsic vacancy site in Ge to achieve the diffusion process [1]. Others like ring mechanism, kick-out mechanism [1][3][5] also plays an important role in diffusion in Ge. In this paper, we apply direct interstitial mechanism to all the elements we calculated to discuss a variety of diffusion behaviors, in which impurities will move exactly along the [111] or equivalent directions between two tetrahedral interstitial sites in Ge lattice [8]. Although it is not true for all the atoms that they diffuse in Ge via direct interstitial mechanism, and the computational condition is set to perform at 0 K which is not the ordinary experimental situation, it is a convenient way to compare all the different diffusion properties and obtain the database to choose proper dopants for desired diffusion behaviors.

## II. COMPUTATIONAL APPROACH

In this work, diffusion energy barriers of 81 elements were calculated using the climbing image nudged elastic band (CI-NEB) method [10], which is implemented in the Vienna ab-initio simulation package (VASP) [11][12]. VTST scripts were used to complete the processing. We define the absolute value of the difference between highest energy site and lowest energy site either at initial tetrahedral site (T) or the medium pathway site (M) as the migration energy of the direct diffusion mechanism. PAW (projector augmented wave)-PBE (Perdew-Burke-Ernzherof) Pseudopotentials was employed. A supercell with 64 Germanium atoms under zero pressure conditions was used, as well as Brillouin-zone sampling with a Monkhorst-Pack mesh grid of 2×2×2 k-points. The kinetic energy cut-off was set to 400 eV. In self-consistent relaxation, the convergence criterion was set to total energy change no more than $1\times 10^{-5}$ eV and forces on the atoms smaller than 0.02 eV/angstrom. All calculations are completed via National Supercomputer Center (NSCC) in Tianjin.

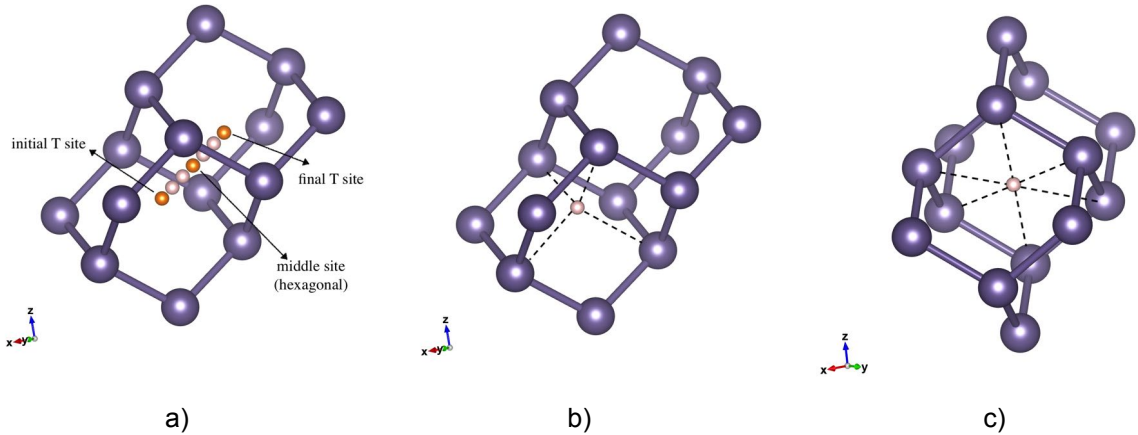

Fig1. a) A schematic path of direct interstitial diffusion in Ge between two nearest tetrahedral sites through the narrowest hexa-coordinate site (T→M→T) b) The T sites, where the small pink sphere indicates the dopant at the initial and final tetrahedral site c) A more clear view of the M site, where the small pink sphere indicates the dopant located at a six-fold coordinate site.

## III. RESULTS AND DISCUSSION

We calculate 81 elements in the periodic table, and all the diffusion barriers along the [111] directions between two nearest T sites are concluded in Fig.1. Extensive experimental studies on the diffusion of typical impurities in Ge have been performed and reported as early as 1954 [13], when macroscopic diffusion behavior in room temperatures in Ge was reported. More recently, the theoretical researches utilizing computational tools gained more attention [5][8][14][16][20]. Our calculation results are proved by those previous researches. The activation enthalpies of impurities doping in Ge for Si, Ge and Sn are arranged in an order as $H_{Si} < H_{Ge} < H_{Sn}$ [1], which is consistent with our results of diffusion energy barriers that $E_{Si}$ (0.303 eV) < $E_{Ge}$ (0.516 eV) < $E_{Sn}$ (0.876 eV), revealing the same diffusion excitation difficulty. For experimental study about another diffusion mechanism in Ge, the vacancy-mediated dopant diffusion [9], the diffusion activation enthalpy was measured as $H_{Sb} < H_{As} < H_P < H_{Sn} < H_{Ga}$, which has the same increasing trend as our results for energy barriers that $E_{Sb}$ (0.305 eV) < $E_{As}$ (0.600 eV) < $E_P$(0.730 eV) < $E_{Sn}$ (0.876 eV) < $E_{Ga}$ (1.087 eV). Additional, experimental applications performed by Brotzmann et al [15] hold for our calculations, too. Brotzmann et al [15] reported that the diffusion depth of As in Ge is shallower than that of Sb in experiment, which means the diffusion of As in Ge is more difficult than Sb. Our calculations are able to explain the phenomenon well that the diffusion barrier of As (0.600 eV) is larger than that of Sb (0.305 eV), while the atomic radius of As is much smaller than Sb's. The early research [17] about deep level impurities in Ge showed a relative quick diffusivity of Cu and Ni, while a much slower diffusivity of Zn and Te, indicating a small diffusion barrier for Cu and Ni and higher barrier for the latter two, Zn and Te. This is confirmed by our calculation, in which the diffusion barriers are Cu (0.089 eV), Ni (0.259 eV), Zn (0.614 eV), Te (0.825eV), respectively. Specifically, two metal atoms, lithium (Li) and copper (Cu), were calculated migration energies in Ge as 0.54 eV and 0.14 eV, respectively [5],which fits well with the previous experimental work showing for Li (0.4 – 0.5 eV) [18] and Cu (0.084 eV) [19]. That perfectly corresponds to our results, for which 0.486 eV for diffusion of Li and 0.089 eV for Cu. Palladium (Pd) that has a large atomic radius (139 pm) but low diffusion barrier (0.033 eV) is also proved by the ultrafast diffusivity of Pd in Ge lattice [5].

It can be clearly seen from the general diffusion barrier database in figure 2 a) and table 1 in the supplementary material that the majority of the elements are in excellent agreement with the size effect, which means the diffusion barrier increases when the atomic radii increase due to the largest strain energy at the narrowest M site, but still a bunch of elements have relatively large errors to the predicted proportionally increasing trend line. Furthermore, some elements have the lowest energy value at M site, which is a totally adverse condition of size dominate diffusion mechanism.

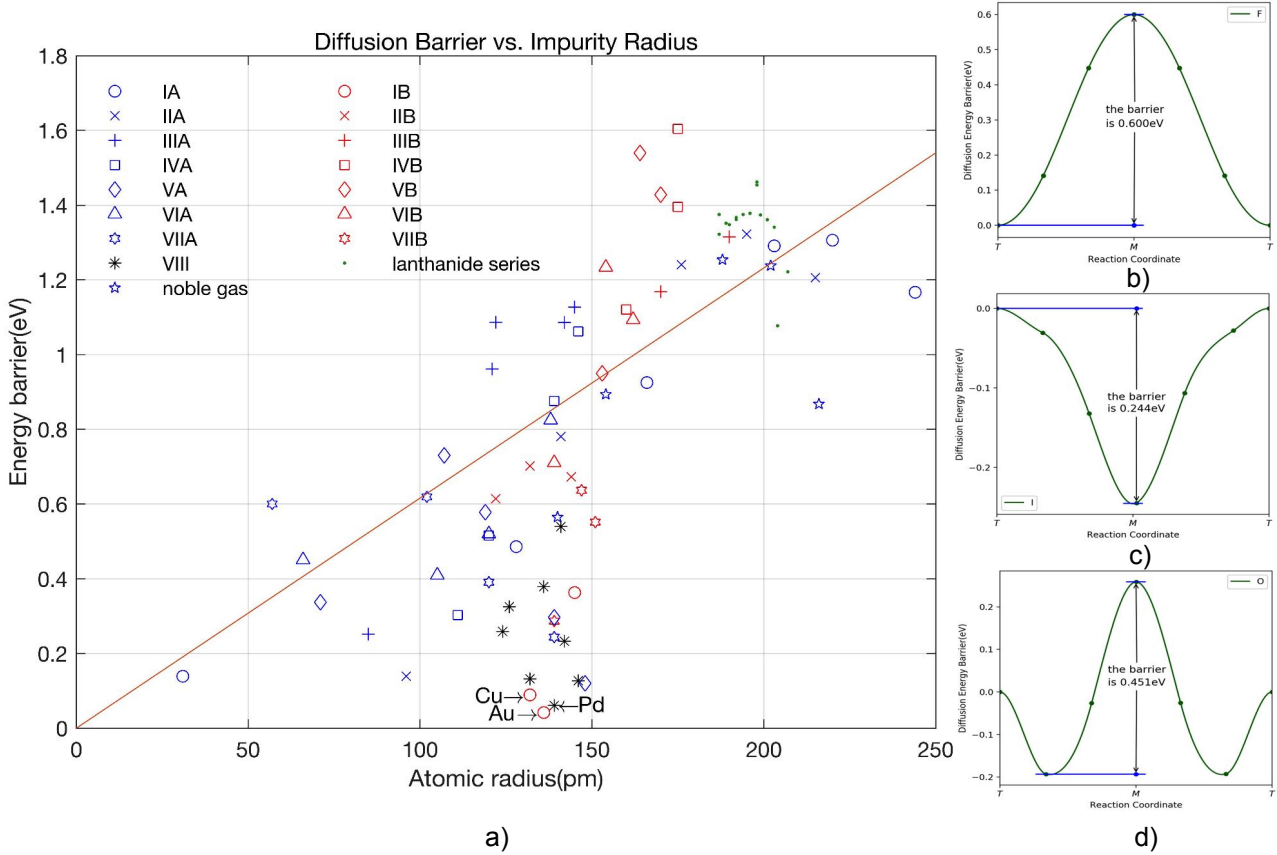

Fig.2. a) The diffusion barriers database of atoms in Ge, including 81 common elements in the periodic table. The barrier is defined as the difference between the highest and lowest energy values among the seven images along the diffusion pathway, which images are evenly inserted between the initial and final T sites based on NEB method. b) c) d) Three typical types of diffusion energy profiles.

There are typically three types of diffusion profiles. Based on the size effect of the dopant atoms [8][16], the diffusion energy profiles between two T sites should have a mountain shape as is shown in figure 2 b), which reaches its maximum energy value at the M site. Because the initial T site has a larger interstitial space than the narrow M site, atom will suffer from greater resistance at M site thus needs to overcome a relative large strain energy there. With the increasing atomic radius in a series of elements, the energy barrier values are also supposed to increase. According to our calculations, IA, IIA, IIA, IVA and VIA groups all correspond to the size dominate rules, which means energy barriers increase as the atomic radius getting larger. That some of the most oversized atoms including Cs and Ba do not have the largest barrier values are probably due to the Ge lattice distortion in the relaxation process, which counteract the increasing strain energy.

Counterintuitively, there are atoms that have adverse shape diffusion energy profiles which have maximum energy at the initial T site, but most stable site at M or somewhere between T and M site (it shows a U shape in figure 2 c) and a W shape in d) ). VA, VIA and VIIA group elements all show this property when they diffuse in Ge. Specifically, the VA group elements have a totally adverse energy barrier decreasing trend with the increasing atomic radius. This leads us to consider other effects contributing to the diffusion behaviors in Ge.

Here we consider the overall diffusion barrier as a result of a combination of bonding effect and size effect. The diffusion energy profile resulting from bonding with Ge lattice have a U shape like figure 2 (c), because the bonding is stronger at the hexagonal M site than it at the initial T site. According to electron localization function (ELF) analysis (see supplementary material) [21], VA group elements (P, As, Sb, Bi) have similar electronic overlap between impurities and Ge lattice atoms. Considering their diffusion barriers should have increased with the increasing atomic radius, the energy barriers decrease above a similar level to a minus value due to bonding. Thus, the final energy barrier which is the absolute value of a

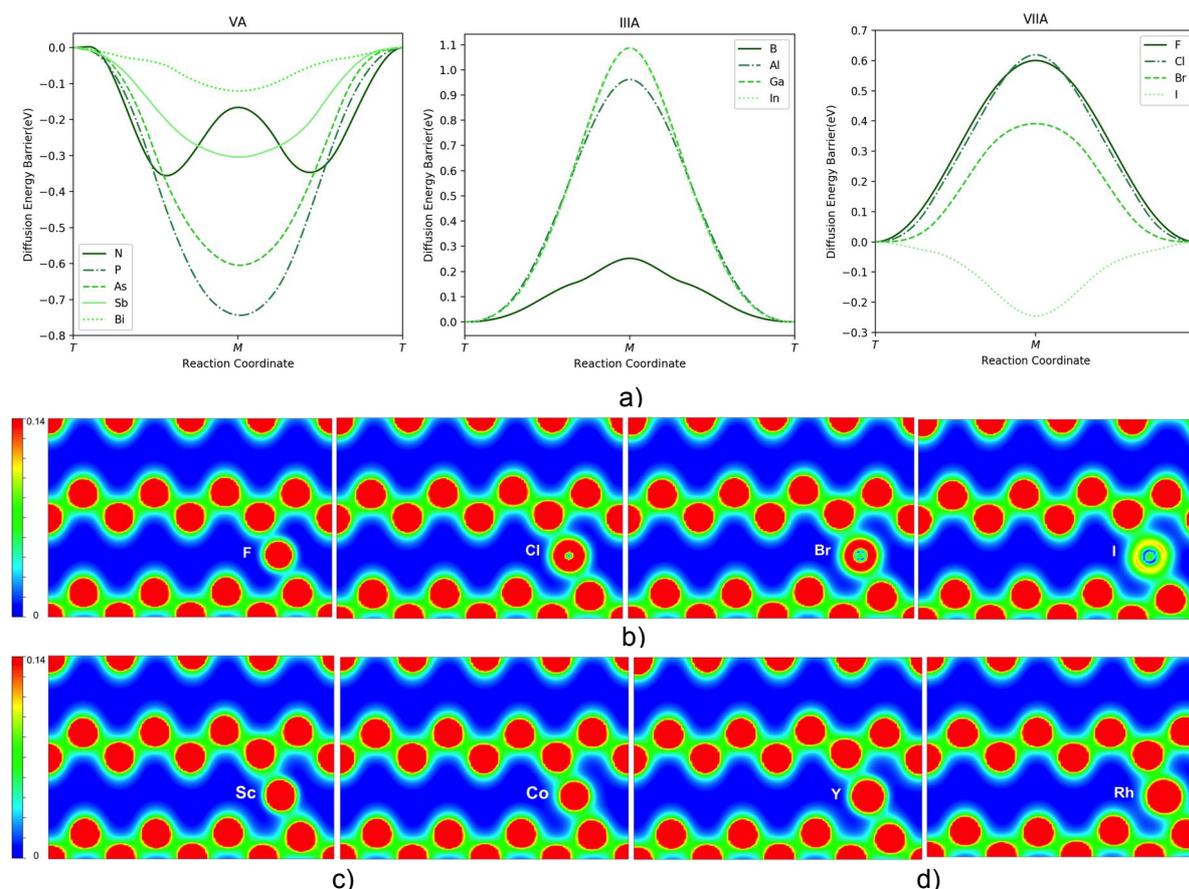

Fig.3 a) Several groups of diffusion barriers. Bonding effect can be observed apparently in the Halogen elements, which theory is then properly applied to IIIA and VA groups. b) The charge density analysis of Halogen elements (F, Cl, Br, I), where increasing bonding can be observed through the electron cloud overlap. c) The charge density analysis of 3d transition metal elements (Sc, Co) and d) 4d transition metal elements (Y, Rh). Increasing bonding is clearly shown in figures.

series of increasing minus values will decrease as the atomic size increases. Similarly, obvious increasing bonding in VIIA Halogen element (F, Cl, Br, I) impurities diffusing in Ge can be observed from ELF analysis. While F atom almost has no bonding with its nearest Ge lattice atoms, larger area electron clouds overlap appears between Cl, Br, I and Ge atoms. Therefore, the bonding gets stronger in this group, and the diffusion energy profile turns to a U shape at the end.

Boron (B), aluminum (Al), gallium (Ga), and indium (In) are acceptors in periodic table that can gain electrons and potentially serve as p-type dopants in Ge-based materials modification and devices fabrication [1][9]. According to our calculations, the diffusion energy barriers basically increase with the increasing atomic sizes. It can be explained by the size effect simply, but the bonding effect can also contribute to this rule. Based on the ELF analysis (see supplementary material), the bonding level of B, Al, Ga and In with Ge during the diffusion process has not changed much. B has relative stronger bonding which leads to a much lower diffusion barrier than Al's. In general, the bonding level is not high enough to reverse the strain-controlled energy curves, so the overall energy values along the pathway still have a convex shape and increase with the atomic radius.

To verify the existence of bonding and quantitively measure the bonding level, we calculated the diffusion barriers of 3d transition metal (TM) atoms with and without lattice relaxation, and compared the difference with their most stable site in Ge lattice[16]. As is shown in table 1, the early 3d TM atoms with large atomic radii show a relatively large difference between the diffusion barriers with and without lattice relaxation, while the late 3d TM atoms including Fe, Co, Ni and Cu show very small change in the diffusion barrier before and after lattice relaxation. This can be attributed to the effect of bonding, because we found those elements that are sensitive to lattice relaxation (have energy barrier changes larger than 0.1eV) all have their most stable sites at T site, while those late 3d TM elements that are not affected by lattice relaxation much have their maximum stability at M site. Cu element remains the most stable site at T position, but with an extremely low energy. This abnormal stability at most narrow M site should be caused by their strong bonding with the nearest Ge atoms, which is at the center of a six-fold-coordinated Ge ring.

According to the pseudo-Jahn-Teller interaction theory [16][22], late 3d TM elements have special valence shell structure as $3d^n4s^0$ (n>6) which has a smaller excitation energy of jumping to $3d^{n-1}4s^1$ than that of early 3d TM elements with small n, resulting in a stronger d-s coupling. Through comparing the charge density and electron localization function of those atoms at M sites, larger area electron cloud overlaps appear at the late 3d TM elements in Ge.

| 3d transition metal element | Sc | Ti | V | Cr | Mn | Fe | Co | Ni | Cu |
|---|---|---|---|---|---|---|---|---|---|
| Atomic radius (pm) | 170 | 160 | 153 | 139 | 139 | 132 | 126 | 124 | 132 |
| TM energy difference - Calculation without lattice relaxation (eV) | 0.254 | 0.703 | 0.702 | 0.533 | 0.137 | 0.178 | 0.368 | 0.301 | 0.089 |
| Calculation with lattice relaxation (eV) | 1.169 | 1.121 | 0.950 | 0.711 | 0.287 | 0.132 | 0.325 | 0.259 | 0.089 |
| Most stable atomic site | T | T | T | T | T | M | M | M | T |

Table 1. Atomic radius, calculated diffusion barriers without lattice relaxation, calculated diffusion barriers with lattice relaxation of 3d transition metal dopants in Ge. The most stable site of the dopant is indicated in the table.

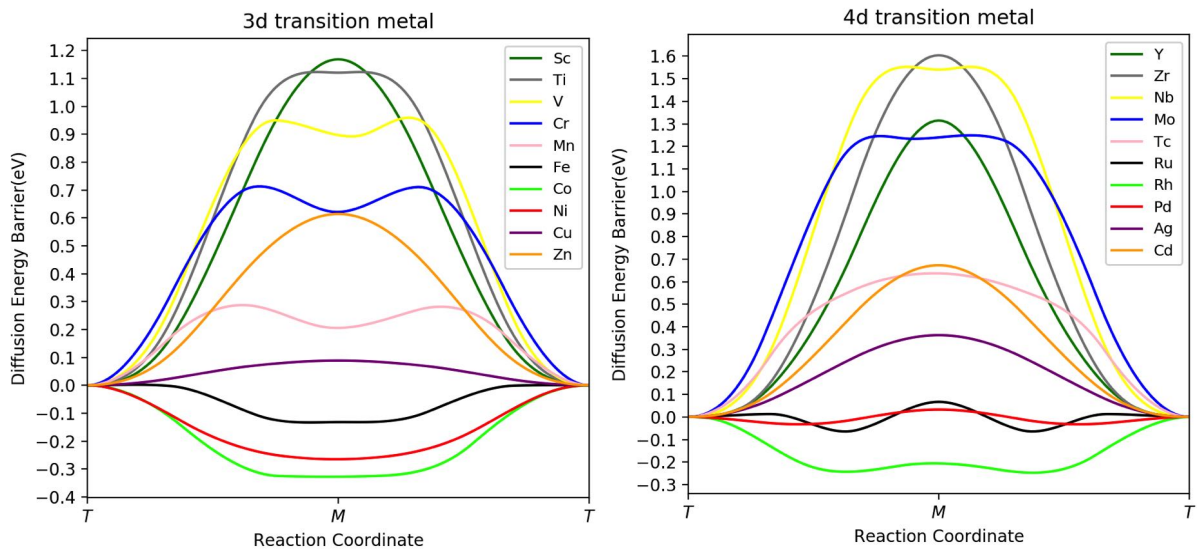

Fig. 4. Diffusion barrier profiles of 3d and 4d transition metal elements. A generally increasing bonding can be inferred from the changing trend of the diffusion energy curves. Five images along the diffusion pathway inserted by NEB method are omitted.

This energy gain of bonding contributes to the total diffusion energy and finally results in the shape reversal of diffusion energy profile from a convex shape (figure 2 b)) to a concave shape (figure 2 c) U shape and d) W shape). Specifically, although Cu has its most stable site at T, its diffusion barrier is extremely small comparing to its relatively large atomic radii. Elements with similar radius, for example, Li (128pm) has a barrier of 0.486eV. This indicates that the strong size effect is too hard to overcome by bonding-stabling effect. Additionally, the profile does turn to a nearly concave shape in calculation without lattice relaxation, which to some extent attests to our speculation. Since 4d TM elements have similar valence electron shell structure as 3d elements, similar increasing bonding trend is supposed to exist there. This conjecture is confirmed by the slightly reversed energy curves of Ru, Rh and Pd, along with the larger and larger electron cloud overlap in charge density analysis and the obvious differences in diffusion barriers before and after lattice relaxation (see supplementary material).

To get a full appreciation of what we have proved in the transition metal elements, we apply the theory to the Halogen elements. With valence shell structure of $ns^2np^5$, the vacant d-orbitals of Cl, Br and I can participate in bonding with small excitation energies, while F atom without chemically active d bands can only form a minus-one compound by filling the p-orbitals. So only Cl, Br and I can have strong coupling between the low-lying excited states and ground states due to pseudo Jahn-Teller effect [16], F has no chance of such s-d coupling. This prediction fits well with the observed no-bonding state for F at M site and the generally stronger bonding in Cl, Br and I there (figure 3 b)).

## IV. SUMMARY

In conclusion, we obtained the direct interstitial diffusion barriers in germanium of 81 elements in the periodic table based on first-principle calculation. For those elements with anomalous diffusion barrier values and energy profiles (which contradict to the general rules dominated by size effect that the diffusion energy is a concave curve and the barrier increases with larger size impurity), we shed light on the possible mechanism as a combination of size effect and bonding effect. The pseudo Jahn-Teller interaction is involved to explain the formation of bonding in the diffusion process. Although it is difficult to completely rule out all the possible mechanisms existed in diffusion of so many elements, we firstly propose the bonding rule for a wide range of elements in Ge, and provide solid evidence of the existence of different levels of bonding. By comparing the diffusion barriers in our database, researchers will be able to better choose proper impurities to fabricate electrodes, heterostructures and transistors for an expected diffusivity. For example, it can be observed apparently from the database that subgroup elements are good sources for ultrafast dopant materials and impurities to make shallow junction, since they include those special elements with extremely low diffusion barriers which are supposed to realize bulk doping in quick diffusivity, and those with small atomic size but high diffusion barriers which are ideal materials for fabricating electrodes that do not want deep level impurities. Therefore, Ge-based semiconductor devices with more desirable properties are promising to be designed and manufactured.

# *Supplementary Material

## I. Table of diffusion barrier database

| AN | IE | AR/pm | DE/eV | DOE | MSS | AN | IE | AR/pm | DE/eV | DOE | MSS |
|---|---|---|---|---|---|---|---|---|---|---|---|
| 1 | **H** | 31 | 0.139 | size | T | 43 | **Tc** | 147 | 0.637 | size | T |
| 2 | **He** | 140 | 0.565 | size | T | 44 | **Ru** | 146 | 0.127 | both | O |
| 3 | **Li** | 128 | 0.486 | size | T | 45 | **Rh** | 142 | 0.233 | both | O |
| 4 | **Be** | 96 | 0.139 | size | T | 46 | **Pd** | 139 | 0.061 | both | O |
| 5 | **B** | 85 | 0.252 | size | T | 47 | **Ag** | 145 | 0.363 | size | T |
| 7 | **N** | 71 | 0.337 | both | O | 48 | **Cd** | 144 | 0.673 | size | T |
| 8 | **O** | 66 | 0.451 | both | O | 49 | **In** | 142 | 1.086 | size | T |
| 9 | **F** | 57 | 0.600 | size | T | 50 | **Sn** | 139 | 0.876 | size | T |
| 10 | **Ne** | 154 | 0.894 | size | T | 51 | **Sb** | 139 | 0.297 | bonding | M |
| 11 | **Na** | 166 | 0.925 | size | T | 52 | **Te** | 138 | 0.825 | bonding | M |
| 12 | **Mg** | 141 | 0.781 | size | T | 53 | **I** | 139 | 0.244 | bonding | M |
| 13 | **Al** | 121 | 0.962 | size | T | 54 | **Xe** | 216 | 0.868 | size | T |
| 14 | **Si** | 111 | 0.303 | size | T | 55 | **Cs** | 244 | 1.167 | size | T |
| 15 | **P** | 107 | 0.730 | bonding | M | 56 | **Ba** | 215 | 1.206 | size | T |
| 16 | **S** | 105 | 0.410 | bonding | M | 57 | **La** | 207 | 1.222 | size | T |
| 17 | **Cl** | 102 | 0.619 | size | T | 58 | **Ce** | 204 | 1.077 | size | T |
| 18 | **Ar** | 188 | 1.254 | size | T | 59 | **Pr** | 203 | 1.341 | size | T |
| 19 | **K** | 203 | 1.291 | size | T | 60 | **Nd** | 201 | 1.361 | size | T |
| 20 | **Ca** | 176 | 1.241 | size | T | 61 | **Pm** | 199 | 1.374 | size | T |
| 21 | **Sc** | 170 | 1.169 | size | T | 62 | **Sm** | 198 | 1.462 | size | T |
| 22 | **Ti** | 160 | 1.121 | size | T | 63 | **Eu** | 198 | 1.454 | size | T |
| 23 | **V** | 153 | 0.950 | size | T | 64 | **Gd** | 196 | 1.378 | size | T |
| 24 | **Cr** | 139 | 0.711 | size | T | 65 | **Tb** | 194 | 1.375 | size | T |
| 25 | **Mn** | 139 | 0.287 | size | T | 66 | **Dy** | 192 | 1.367 | size | T |
| 26 | **Fe** | 132 | 0.132 | bonding | M | 67 | **Ho** | 192 | 1.362 | size | T |
| 27 | **Co** | 126 | 0.325 | bonding | M | 68 | **Er** | 189 | 1.352 | size | T |
| 28 | **Ni** | 124 | 0.259 | bonding | M | 69 | **Tm** | 190 | 1.348 | size | T |
| 29 | **Cu** | 132 | 0.089 | size | T | 70 | **Yb** | 187 | 1.375 | size | T |
| 30 | **Zn** | 122 | 0.614 | size | T | 71 | **Lu** | 187 | 1.322 | size | T |
| 31 | **Ga** | 122 | 1.087 | size | T | 72 | **Hf** | 175 | 1.395 | size | T |
| 32 | **Ge** | 120 | 0.516 | size | T | 73 | **Ta** | 170 | 1.428 | size | T |
| 33 | **As** | 119 | 0.578 | bonding | M | 74 | **W** | 162 | 1.094 | size | T |

| 34 | Se | 120 | 0.520 | bonding | M | 75 | Re | 151 | 0.552 | size | T |
| 35 | Br | 120 | 0.391 | size | T | 77 | Ir | 141 | 0.540 | bonding | M |
| 36 | Kr | 202 | 1.238 | size | T | 78 | Pt | 136 | 0.379 | bonding | M |
| 37 | Rb | 220 | 1.306 | size | T | 79 | Au | 136 | 0.042 | size | T |
| 38 | Sr | 195 | 1.323 | size | T | 80 | Hg | 132 | 0.702 | size | T |
| 39 | Y | 190 | 1.315 | size | T | 81 | Tl | 145 | 1.127 | size | T |
| 40 | Zr | 175 | 1.604 | size | T | 82 | Pb | 146 | 1.062 | size | T |
| 41 | Nb | 164 | 1.540 | size | T | 83 | Bi | 148 | 0.120 | bonding | M |
| 42 | Mo | 154 | 1.234 | size | T | | | | | | |

Table 1. Summary of calculated results: AN means atomic number, IE means impurity element, R means atomic radius (pm), DE means diffusion barrier (eV), DOE means dominate effect, MSS means the most stable site among NEB images. (T – tetrahedral site, M - medium site, O – other images except T and M)

## II. Electron localization function analysis

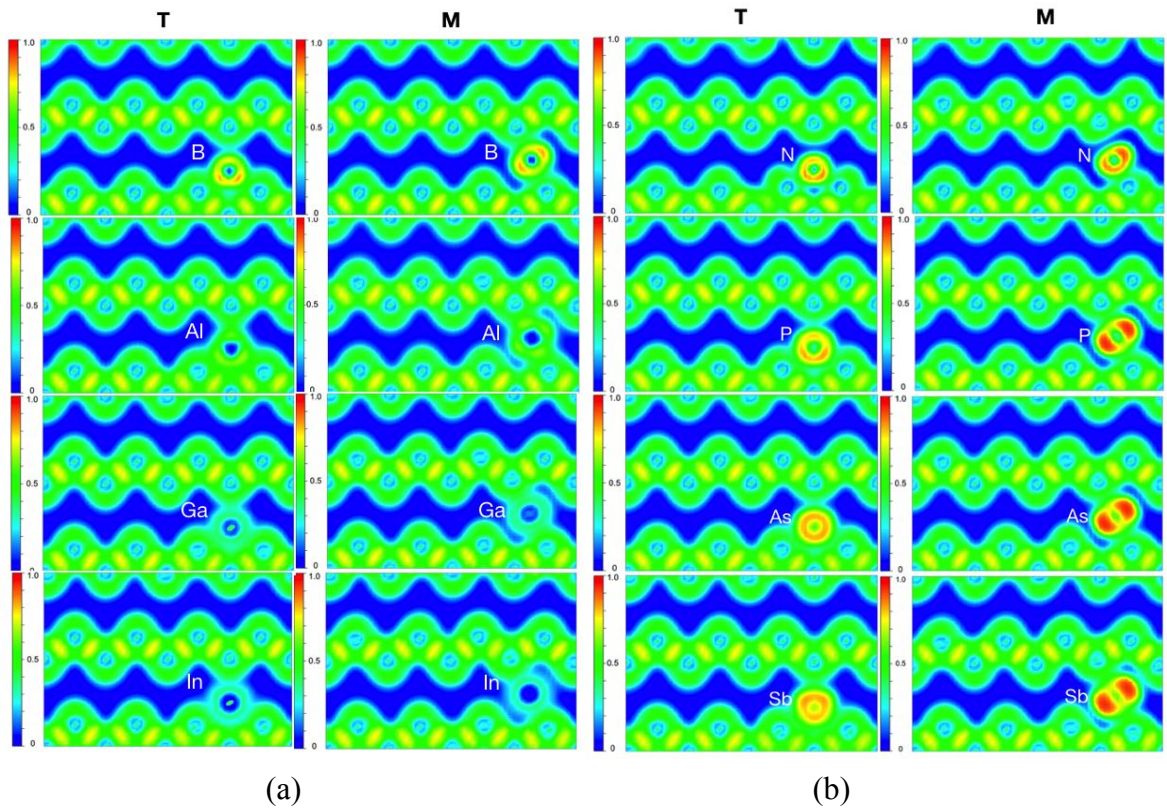

(a)            (b)

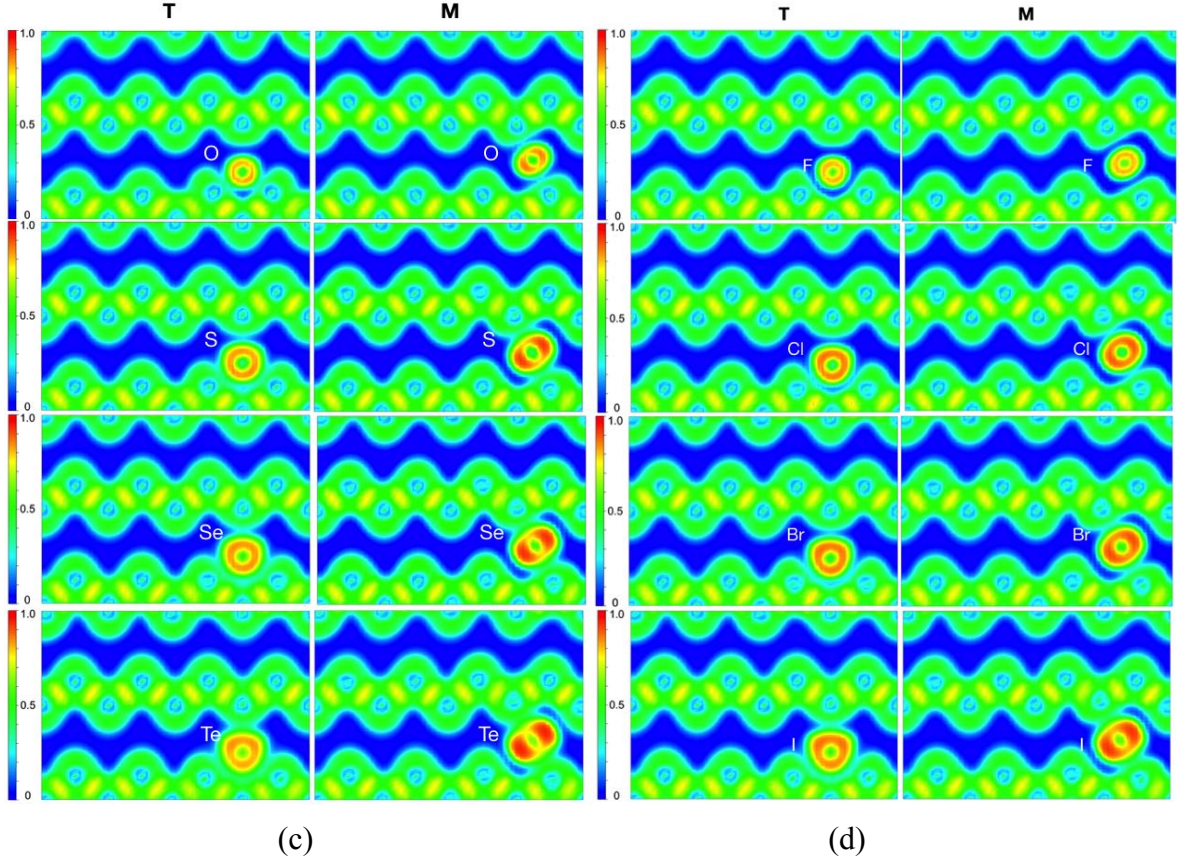

(c)                                    (d)

Fig 1. a) The electron localization function analysis of IIIA elements(B, Al, Ga and In), left is the tetrahedral site and right is the medium site b) The electron localization function analysis of VA elements(N, P, As and Sb) c) The electron localization function analysis of VIA elements(O, S, Se, Te) d) The electron localization function analysis of VIIA elements(F, Cl, Br and I)

### III. Calculated diffusion barrier with and without lattice relaxation

| 4d transition metal element | Y | Zr | Nb | Mo | Tc | Ru | Rh | Pd | Ag |
|---|---|---|---|---|---|---|---|---|---|
| Atomic radius (pm) | 190 | 175 | 164 | 154 | 147 | 146 | 142 | 139 | 145 |
| TM energy difference - Calculation without lattice relaxation (eV) | 0.758 | 0.279 | 0.918 | 0.860 | 0.314 | 0.202 | 0.483 | 0.332 | 0.416 |
| Calculation with lattice relaxation (eV) | 1.315 | 1.604 | 1.540 | 1.234 | 0.637 | 0.067 | 0.233 | 0.033 | 0.363 |
| Most stable atomic site | T | T | T | T | T | btTM | btTM | btTM | T |

Table 2. Atomic radius, calculated diffusion barriers without lattice relaxation, calculated diffusion barriers with lattice relaxation of 3d transition metal dopants in Ge. The most stable site of the dopant is indicated in the table.